\documentstyle[a4,epsf,12pt]{article}

\newcommand{\cL}{{\cal L}}
\newcommand{\cO}{{\cal O}}
\newcommand{\Lmd}{\Lambda}
\renewcommand{\Re}{{\rm Re}}
\renewcommand{\Im}{{\rm Im}}
\newcommand{\hL}{h_{\rm L}}
\newcommand{\hR}{h_{\rm R}}
\newcommand{\df}{{\rm d}}
\newcommand{\del}{\partial}
\newcommand{\Acl}{A_{\rm cl}}
\newcommand{\Iinf}{I_{\infty}}
\newcommand{\lmdL}{\lambda_{\rm L}}
\newcommand{\lmdR}{\lambda_{\rm R}}
\newcommand{\lmdl}[1]{\lambda_{{\rm L}#1}}
\newcommand{\lmdr}[1]{\lambda_{{\rm R}#1}}
\newcommand{\chiL}{\chi_{\rm L}}
\newcommand{\chiR}{\chi_{\rm R}}
\newcommand{\chil}[1]{\chi_{{\rm L}#1}}
\newcommand{\chir}[1]{\chi_{{\rm R}#1}}
\newcommand{\hl}[1]{h_{{\rm L}#1}}
\newcommand{\hr}[1]{h_{{\rm R}#1}}
\newcommand{\lmd}{\lambda}
\newcommand{\yefij}{y_{{\rm eff}ij}}
\newcommand{\Ml}[1]{M_{{\rm L}#1}}
\newcommand{\Minf}[1]{M_{\infty #1}}
\newcommand{\Mninf}[1]{M_{-\infty #1}}

\begin{document}

\begin{titlepage}
\null
\begin{flushright}
 {\tt hep-th/0011098}\\
TIT/HEP-459
\\
November 2000
\end{flushright}

\vskip 2cm
\begin{center}
{\LARGE \bf  Our wall as the origin of CP violation}

\lineskip .75em
\vskip 2.5cm

\normalsize

  {\large \bf Yutaka Sakamura}
{\def\thefootnote{\fnsymbol{footnote}}
\footnote[5]{\it  e-mail address:
sakamura@th.phys.titech.ac.jp}}

\vskip 1.5em

{\it Department of Physics, Tokyo Institute of Technology\\
Tokyo 152-8551, JAPAN }

\vspace{18mm}

{\bf Abstract}\\[5mm]
{\parbox{13cm}{\hspace{5mm} \small
A possibility that the origin of the CP violation in our world is 
a domain wall itself, on which we are living, is investigated
in the context of the brane world scenario.
We estimate the amount of the CP violation
on our wall, and show that either of order one or suppressed CP phases 
can be realized.
An interesting case where CP is violated due to the {\it coexistence} of 
the walls, which conserve CP individually, is also considered.
We also propose a useful approximation for the estimation of 
the CP violation in the double-wall background.
}}

\end{center}

\end{titlepage}

\clearpage

\section{Introduction}
Recently, much attention is paid to topological objects such as D-branes
in string theories\cite{polchinski}, BPS domain walls\cite{dvali} and 
junctions\cite{gibbons} in supersymmetric field theories.
In particular,  the so-called ``brane world '' scenario, 
in which our four-dimensional world on these topological objects is embedded 
into a higher dimensional space-time, is investigated actively.
In such scenarios, the standard model fields are confined to the brane, 
whereas the gravity propagates in the bulk space-time.

The authors of Refs.~\cite{arkani,antoni} pointed out that 
the hierarchy problem between the Planck and weak scales 
may be addressed to the existence of some large extra dimensions.
If such extra dimensions exist, the Planck scale $M_{\rm pl}$ of our world 
is not fundamental
and related to the genuine fundamental scale $M_{\ast}$ of a higher 
dimensional theory
by $M_{\rm pl}^{2}\sim M_{\ast}^{2+n}V_{n}$, where $n$ and $V_{n}$ are 
a dimension and a volume of the compact extra space respectively.
Thus $M_{\ast}$ can be the TeV scale by supposing that the radii of
the compactified extra dimensions are large compared to $1/M_{\rm pl}$.
However there is still a large hierarchy between $M_{\ast}$ and 
$1/(V_{n})^{1/n}$ in this scenario.

In contrast, a rather different idea was proposed to solve the hierarchy 
problem in Ref.~\cite{randall}.
Their model consists of our four-dimensional world and another three-brane, 
which are located at fixed points of an orbifold $S^{1}/Z_{2}$ whose radius
is $r_{c}$.
Assuming a non-factorizable metric of the bulk space-time, the weak scale
$M_{\rm w}$ is related to the fundamental scale $M_{\ast}$, which is of order
$M_{\rm pl}$ in this scenario, 
by $M_{\rm w}=e^{-\pi k r_{c}}M_{\ast}$,
where $k$ is a parameter of order $M_{\ast}$.
For the sake of the exponential dependence on the radius $r_{c}$, 
we can obtain large hierarchy between $M_{\rm w}$ and $M_{\rm pl}$
without introducing any hierarchy among the fundamental parameters.
Indeed, if $kr_{c}\sim 12$, the observed hierarchy $M_{\rm pl}/M_{\rm w}\sim
10^{16}$ is realized.

In addition, it was pointed out in Ref.~\cite{dvali2,kaplan} that 
the hierarchy 
among the fermion masses in the standard model may be explained by localizing 
fermions in different generations on different positions in 
the extra dimension.
In their scenario, the fermion mass hierarchy is realized 
by the coupling to a Higgs condensate that falls off exponentially away from 
the position where the Higgs is localized.

Many other applications of the brane world scenario to cosmology and 
astrophysics are also investigated\cite{maartens,shiromizu}.

On the other hand, the origin of the CP violation is still a mystery
in particle physics.
One of its candidates is the spontaneous CP violation\cite{lee}.
This is convincing as a solution to the strong CP problem\cite{holman,choi}
and can easily control CP violating phases appearing in the supersymmetric 
(SUSY) standard models or 
models with enlarged Higgs sector\cite{babu,masip,liu}.
This scenario is based on the assumption that CP is an exact symmetry 
in high energy region but violated by the complex vacuum expectation values 
(VEVs) of some scalar fields at low energy.

In the brane world scenario, there is another possibility of 
the spontaneous CP violation.
Namely, CP is conserved in the fundamental bulk theory but violated by
the {\it background field configuration} 
instead of the complex VEVs of the scalar fields on our wall.
We will investigate such a case in this paper.

Here we will assume that we are living on a four-dimensional domain wall, 
which interpolates two degenerate vacua of a five-dimensional bulk theory.
We will call this domain wall ``our wall'' in this paper.
In general, such a domain-wall field configuration can have 
a non-trivial complex phase even if
parameters of the bulk theory are all real.

The purpose of this paper is to propose a mechanism of the CP violation
induced by our wall itself.
We investigate how the CP violation appears in the effective theory 
and estimate the magnitude of the typical CP phases.
Then we show our mechanism can be used to realize large class of 
models with realistic CP violation.

In this paper, we do not discuss any gravitational effects nor gauge fields
for simplicity.
However, the qualitative features discussed here are supposed to be general 
and does not depend on the details of the theory.

The paper is organized as follows.
In the next section we will explain our mechanism and give a few examples
of the ``complex domain-wall configurations''.
Then in Section~\ref{estimation} we will introduce matter fields, which are
trapped on our wall, and estimate the CP violation in the four-dimensional
effective theory.
In Section~\ref{coexist}, another interesting model is considered.
This model has two kinds of domain walls which conserve CP individually, 
but violate it when the two walls coexist.
Section~\ref{summary} is devoted to the summary.
In the appendix, we introduce a particular model that has 
a calculable double-wall configuration, 
and confirm the validity of the approximation we used in Section~\ref{coexist}.

\section{CP violation due to the wall}
In this section, we will illustrate our CP-violating mechanism by
introducing a few simple toy models.
At first, we will explain the meaning of the ``complex configuration''.

Let us consider the following five-dimensional scalar theory.

\begin{equation}
 \cL=-\del^{M}A^{\ast}\del_{M}A-\Lmd^{5}\left|
 1-\frac{g^{n}}{\Lmd^{n}}A^{n}\right|^{2}, \;\;\;(M=0,1,\cdots,4)
 \label{phi-n-th}
\end{equation}
where $\Lmd$ is a parameter that has a dimension of the mass and $g$ is 
a coupling constant whose mass-dimension is $-1/2$, and both of them are
assumed to be real and positive.
This model has $n$ degenerate vacua: $A=\Lmd/g,(\Lmd/g)e^{2\pi i/n},
(\Lmd/g)e^{4\pi i/n},\cdots, (\Lmd/g)e^{2(n-1)\pi i/n}$.
There is a domain-wall configuration interpolating the adjacent vacua
$A=\Lmd/g$ and $(\Lmd/g)e^{2\pi i/n}$.

The most simple case is that of $n=2$. 
In this case, both vacua $A=\pm\Lmd/g$ lie on the real axis of the complex
target space, and the domain-wall configuration is a well-known kink
solution,
\begin{equation}
 \Acl(y)=\frac{\Lmd}{g}\tanh\{g\Lmd^{3/2}(y-y_{0})\},
\end{equation}
where $y$ is a coordinate of the fifth dimension, i.e. $y\equiv x^{4}$, and 
$y_{0}$ is a location of the domain wall.
Note that this is a real function and thus has no complex phase.

\begin{figure}
\leavevmode
\epsfysize=5cm
\centerline{\epsfbox{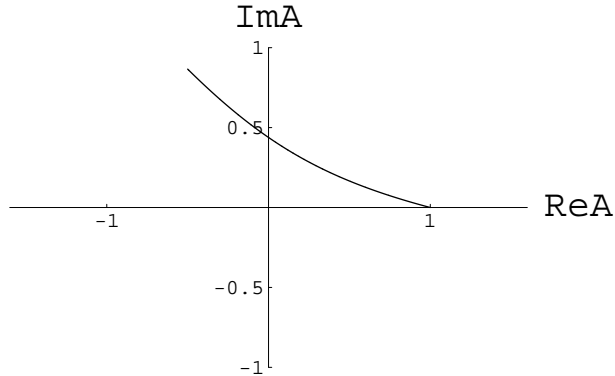}}
\caption{An example of the complex configuration. This is a
domain-wall configuration interpolating the vacua: $A=0$ and
$(\Lmd/g)e^{2\pi i/3}$, of the model Eq.(\ref{phi-n-th}) 
in the case of $n=3$. 
The field $A$ is normalized by $\Lmd/g$ in the figure.}
\label{simple-ex}
\end{figure}

When $n$ is greater than two, however, $(\Lmd/g)e^{2\pi i/n}$ becomes
a complex number and the field configuration under consideration
no longer lies on the real axis, as shown in Fig.\ref{simple-ex}.
In other words, this background configuration has a non-trivial complex
phase that depends on the coordinate of the fifth dimension $y$.
We will call such a configuration a ``complex configuration''.

For another example of the complex configuration, 
we can take a domain wall proposed in
Ref.~\cite{hofmann}.
Their model is\footnote{
In Ref.~\cite{hofmann}, they discuss a BPS domain wall in 
a four-dimensional ${\cal N}=1$ SUSY theory and parameters
$\Lmd$ and $g$ above are set to be one.}
\begin{equation}
 \cL=-\del^{M}A^{\ast}\del_{M}A-\frac{\Lmd^{8}}{|A|^{2}}
 \left| 1-\frac{g^{2}}{\Lmd^{2}}A^{2}\right|^{2}, \label{comp-ex-model}
\end{equation}
where parameters $\Lmd$ and $g$ have the same mass-dimensions as those of 
the previous model, and are real and positive.

\begin{figure}
\leavevmode
\epsfysize=5cm
\centerline{\epsfbox{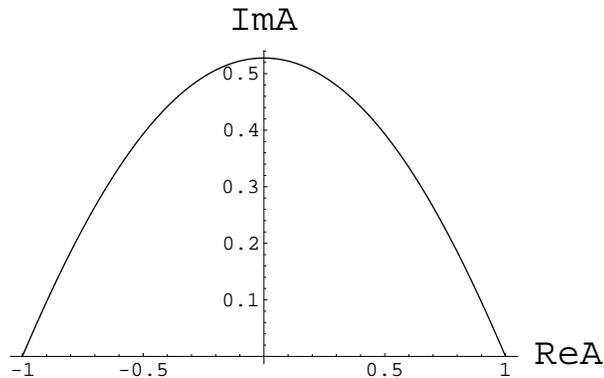}}
\caption{Another example of the complex configuration. This is a
domain wall interpolating the vacua: $A=-\Lmd/g$ and $\Lmd/g$, 
of the model Eq.(\ref{comp-ex-model}).
The field $A$ is normalized by $\Lmd/g$.}
\label{complex-ex}
\end{figure}

This theory has two vacua $A=\pm\Lmd/g$.
Though these vacua are both on the real axis,
the domain wall interpolating them becomes a complex configuration
due to the singularity of the potential at the origin. 
(See Fig.\ref{complex-ex}.)

It should be noted that five-dimensional theories are non-renormalizable
and have physical meanings only as effective theories of some more 
fundamental theories.
So there is no reason for thinking highly non-renormalizable theory
like Eq.(\ref{phi-n-th}) or Eq.(\ref{comp-ex-model}) to be unnatural.

The $y$-dependent complex phases that these walls have will become
the source of the CP violation in the low-energy theory.

In the following, we will assume that our wall 
is a complex configuration and discuss the CP violation that appear
in the low-energy effective theory.

\section{Effective theory and CP violation} \label{estimation}
In this section, we will introduce matter fields and investigate 
the zero-modes trapped on our wall, and then estimate the amount
of the observed CP violation in the effective theory of those 
zero-modes.

\subsection{Trapped zero-modes on our wall}
\begin{figure}
\leavevmode
\epsfysize=5cm
\centerline{\epsfbox{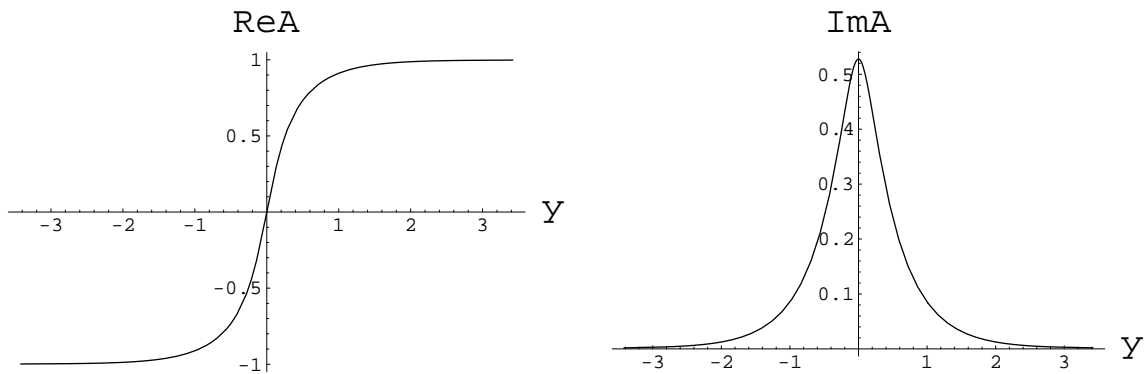}}
\caption{The profile of the domain-wall configuration $\Acl(y)$
of the model Eq.(\ref{comp-ex-model}).
The field $A$ and the coordinate $y$ are normalized by $\Lmd/g$ and
$1/(g^{2}\Lmd^{2})$ respectively.}
\label{wall-profile}
\end{figure}

Here we will take the domain wall of the model 
Eq.(\ref{comp-ex-model}) as an example of the complex domain wall 
identified with our wall, and denote its field configuration as $\Acl(y)$.
The profile of $\Acl(y)$ is shown in Fig.\ref{wall-profile}.
We emphasize that the discussion in this section is completely general
and can be applied to any kind of complex configuration,
as long as $\Re\Acl(y)$ takes different values at $y=\pm\infty$.

At first, we introduce a five-dimensional matter fermion,
\begin{equation}
 \lambda=\left(\begin{array}{c} \lmdL \\ \lmdR \end{array}\right),
\end{equation}
and assume that it interacts with the wall scalar field $A$ as
\begin{equation}
 {\cal L}_{{\rm int}\lambda}=-\frac{\hL}{2}A\bar{\lambda}{\lambda}+h.c.
 =-\hL(\Re A)\bar{\lambda}\lambda, \label{Llmd}
\end{equation}
where the coupling constant $\hL$ is real positive and $\bar{\lambda}$ 
represents the Dirac conjugate of $\lambda$.
Then the linearized equation of motion for $\lambda$ is
\begin{equation}
 -i\gamma^{\mu}\partial_{\mu}\left(\begin{array}{c} \lmdL \\ \lmdR \end{array}
 \right)+\left(\begin{array}{cc} -1 & 0 \\ 0 & 1 \end{array}\right)\partial_{y}
 \left(\begin{array}{c} \lmdL \\ \lmdR \end{array}\right)
 -\hL\Re\Acl\left(\begin{array}{c} \lmdL \\ \lmdR \end{array}\right)=0,
\end{equation}
where $\gamma^{\mu}$ denotes the four-dimensional $\gamma$-matrix
in the chiral representation.
Here by introducing operators 
${\cal O}_{\lmdL}\equiv -\partial_{y}-\hL\Re\Acl(y)$
and ${\cal O}_{\lmdR}\equiv \partial_{y}-\hL\Re\Acl(y)$, 
mode functions $\varphi_{\lmdL,n}(y)$ and $\varphi_{\lmdR,n}(y)$
are defined as solutions of the following mode equations.
\begin{equation}
 {\cal O}_{\lmdL}\varphi_{\lmdL,n}(y)=m_{n}\varphi_{\lmdR,n}(y)\;,\;\;\;
 {\cal O}_{\lmdR}\varphi_{\lmdR,n}(y)=m_{n}\varphi_{\lmdL,n}(y).
\end{equation}

Using these mode functions, the five-dimensional spinor fields can be
expanded as
\begin{equation}
 \lmdL(x,y)=\sum_{n}\varphi_{\lmdL,n}(y)\lmdl{n}(x)\;,\;\;\;
 \lmdR(x,y)=\sum_{n}\varphi_{\lmdR,n}(y)\lmdr{n}(x),
\end{equation}
where $x$ denotes the four-dimensional coordinates.
In this case, the expansion coefficients $\lmdl{n}(x)$ and $\lmdr{n}(x)$ are
regarded as the left- and right-handed components of the four-dimensional 
Dirac spinor fields with masses $m_{n}$ \cite{dvali2}.

For the sake of the index theorem\cite{weinberg}, 
there is a zero-mode in $\lmdL$.
Its mode function $\varphi_{\lmdL,0}(y)$ satisfies an equation 
${\cal O}_{\lmdL}\varphi_{\lmdL,0}(y)=0$,\footnote{
A solution of the equation: ${\cal O}_{\lmdR}\varphi_{\lmdR,0}(y)=0$
is not normalizable.} 
and thus it is
\begin{equation}
 \varphi_{\lmdL,0}(y)=C_{\lmdL}e^{-\hL\int_{0}^{y}\df y'\Re\Acl(y')}\;,
 \label{fr-zero-mode}
\end{equation}
where $C_{\lmdL}$ is a normalization factor.
This mode is localized on our wall ($y=0$)\cite{rubakov}.
Here the corresponding four-dimensional field $\lmdl{0}(x)$ is 
a massless chiral fermion.

Now we will introduce another bulk fermion,
\begin{equation}
 \chi=\left(\begin{array}{c} \chiL \\ \chiR \end{array}\right),
\end{equation}
and assume an interaction with $A$ as
\begin{equation}
 {\cal L}_{{\rm int}\chi}=\hR(\Re A)\bar{\chi}\chi,
\end{equation}
where $\hR$ is real positive.
In the same way, there is a zero-mode localized on our wall,
and its mode function is
\begin{equation}
 \varphi_{\chiR,0}(y)=C_{\chiR}e^{-\hR\int_{0}^{y}\df y'\Re\Acl(y')},
\end{equation}
where $C_{\chiR}$ is a normalization factor.
Note that the corresponding massless field $\chir{0}(x)$ has
an opposite chirality to $\lmdl{0}(x)$.

Next we will investigate a scalar mode trapped on our wall.
At first, let us consider the fluctuation mode around the background
$\Acl(y)$.
\begin{equation}
 A(x,y)=\Acl(y)+\tilde{A}(x,y).
\end{equation}

The linearized equation of motion for $\tilde{A}$ is 
\begin{equation}
 \del^{M}\del_{M}\tilde{A}-\left.\frac{\del f^{\ast}}{\del A^{\ast}}
 \frac{\del f}{\del A}\right|_{A=\Acl}\tilde{A}
 -\left.\frac{\del^{2}f^{\ast}}{\del A^{\ast 2}}f
 \right|_{A=\Acl}\tilde{A}^{\ast}=0, \label{LEOM-A}
\end{equation}
where 
\begin{equation}
 f(A)\equiv\frac{\Lmd^{4}}{A}\left(1-\frac{g^{2}}{\Lmd^{2}}A^{2}\right).
\end{equation}

If we define the operator:
\begin{equation}
 {\cal O}_{A}\equiv -\del_{y}^{2}+\left.\frac{\del f^{\ast}}{\del A^{\ast}}
 \frac{\del f}{\del A}\right|_{A=\Acl}
 +\left.\frac{\del^{2}f^{\ast}}{\del A^{\ast 2}}f\right|_{A=\Acl},
\end{equation}
the mode functions $\phi_{A,n}(y)$ are defined as a solution of
the equation,
\begin{equation}
 {\cal O}_{A}\phi_{A,n}(y)=m_{n}^{2}\phi_{A,n}(y). \label{OA-eq}
\end{equation}

Using these mode functions, we can expand the fluctuation field $\tilde{A}$
as
\begin{equation}
 \tilde{A}(x,y)=\sum_{n}\phi_{A,n}(y)a_{n}(x).
\end{equation}
Here $a_{n}(x)$ is regarded as four-dimensional scalar field with
mass $m_{n}$.

In general, there is a zero-mode in Eq.(\ref{OA-eq}) corresponding to
the Nambu-Goldstone mode (NG mode) for the breaking of 
the translational invariance along the $y$-direction.
Its mode function is 
\begin{equation}
 \phi_{A,0}(y)=C_{A}\del_{y}\Acl(y),
\end{equation}
where $C_{A}$ is a normalization factor.
This mode is localized on our wall.

Finally, we will introduce a bulk complex scalar field $B$.
To obtain a zero-mode localized on our wall, 
we will assume an interaction as follows.
\begin{equation}
 {\cal L}_{{\rm int}B}=-\frac{1}{2}\left(f^{\ast}(A^{\ast})\frac{\del^{2}f}
 {\del A^{2}}(A)B^{2}+h.c.\right)
 -\left|\frac{\del f}{\del A}(A)B\right|^{2}. \label{LB}
\end{equation}

Then the linearized equation of motion for $B$ is
\begin{equation}
 \partial^{M}\partial_{M}B-\left.\frac{\del f^{\ast}}
 {\del A^{\ast}}\frac{\del f}{\del A}\right|_{A=\Acl}B
 -\left.\frac{\del^{2}f^{\ast}}{\del A^{\ast 2}}f\right|_{A=\Acl}B^{\ast}=0. 
 \label{LEOM-B}
\end{equation}
This equation is the same as that for $\tilde{A}(x,y)$ in Eq.(\ref{LEOM-A}), 
and thus there exists a zero-mode in $B$ \footnote{
When we include an additional scalar field like $B$ into the theory,
we should recalculate the classical background, involving all scalar fields.
In our case, however, the field configuration: $A=\Acl$, $B=0$, is still
a minimal-energy configuration (at least at the classical level)
under the boundary condition: $A(y=\pm\infty)=\pm\Lmd/g$, 
$B(y=\pm\infty)=0$, 
although there may be other configuration with the same energy.}and 
its mode function $\phi_{B,0}(y)$ is the same as $\phi_{A,0}(y)$,
that is,
\begin{equation}
 \phi_{B,0}(y)=C_{A}\partial_{y}\Acl(y). \label{phiB0}
\end{equation}
This mode is localized on our wall.

Note that $C_{A}$ is a real number in contrast to $C_{\lmdL}$ and 
$C_{\chiR}$, which are in general complex.
This stems from the fact that 
the linearized equation of motion Eq.(\ref{LEOM-B}) contains both
$B$ and $B^{\ast}$, while that for $\lambda$ or $\chi$ involves only
$\lmd$ or $\chi$ respectively.

\subsection{Estimation of the observable CP violation} \label{est-ob-cp}
In order to discuss a four-dimensional effective theory, we will add 
the following interaction terms to the original Lagrangian 
Eq.(\ref{comp-ex-model}).
\begin{eqnarray}
 \cL_{\rm int}&=&\sum_{i=1}^{n_{g}}\left(M_{\lmd i}\bar{\lmd}_{i}\lmd_{i}
 -\hl{i}(\Re A)\bar{\lmd}_{i}\lmd_{i}\right)
 +\sum_{j=1}^{n_{g}}\left(M_{\chi j}\bar{\chi}_{j}\chi_{j}
 +\hr{j}(\Re A)\bar{\chi}_{j}\chi_{j}\right) \nonumber \\
 &&-\frac{1}{2}
 \left(f^{\ast}(A^{\ast})\frac{\del^{2}f}{\del A^{2}}(A)B^{2}+h.c.\right)
 -\left|\frac{\del f}{\del A}(A)B\right|^{2} \nonumber \\
 &&+\left(\sum_{i,j}y_{ij}B\bar{\chi_{j}}\lambda_{i}+h.c.\right),
 \label{Lint1}
\end{eqnarray}
where $\hl{i},\hr{j}>0$ and $M_{\lmd i}$, $M_{\chi j}$ and $y_{ij}$ are real.
Here $n_{g}$ denotes the number of generations.
The mass terms for the fermions play a role of shifting 
the positions of localized zero-modes to realize the hierarchy among 
the Yukawa couplings\cite{kaplan}.

There are zero-modes $\lmdl{i,0}(x)$, $\chir{j,0}(x)$ and $b_{0}(x)$
in $\lambda_{i}$, $\chi_{j}$ and $B$ respectively, which are all localized 
on our wall.
Their mode functions are 
\begin{eqnarray}
 \varphi_{\lmdL i,0}(y)&=&C_{\lmdL i}e^{-\int_{0}^{y}\df y'
 (\hl{i}\Re\Acl(y')-M_{\lmd i})},
 \nonumber \\
 \varphi_{\chiR j,0}(y)&=&C_{\chiR j}e^{-\int_{0}^{y}\df y'
 (\hr{j}\Re\Acl(y')+M_{\chi j})},
 \nonumber \\
 \phi_{B,0}(y)&=&C_{A}\del_{y}\Acl(y), \label{pNG}
\end{eqnarray}
where $C_{\lmdL i}, C_{\chiR j}$ are complex and $C_{A}$ is real.
To localize fermionic zero-modes $\lmdl{i,0}(x)$ on our wall, 
the bulk mass parameters $M_{\lmd i}$ and couplings $\hl{i}$ 
must satisfy the condition that
functions $f(y)\equiv\hl{i}\Re\Acl(y)+M_{\lmd i}$ should take opposite sign 
at $y=\infty$ and $y=-\infty$ \cite{weinberg}.
The similar condition exists for $M_{\chi j}$ and $\hr{j}$.
We will suppose that these conditions are satisfied.

By integrating out the massive modes, we can obtain a four-dimensional
effective theory of massless zero-modes.
The effective Yukawa couplings $\yefij$ involving $\lmdl{i,0}$, $\chir{j,0}$
and $b_{0}$ are calculated as\cite{dvali,maru} 
\begin{equation}
 \yefij=y_{ij}\int_{-\infty}^{\infty}\df y\phi_{B,0}(y)\varphi_{\chiR j,0}(y)
 \varphi_{\lmdL i,0}(y). \label{yeff}
\end{equation}

The CP violation in this effective theory appears as the complex phases of
the Yukawa couplings, i.e. \ $\zeta_{ij}\equiv\arg(\yefij)$.
However some of them can be absorbed by the field redefinition of
$\lmdl{i,0}(x)$ and $\chir{j,0}(x)$.
The number of the Yukawa couplings is $n_{g}^{2}$ and all the couplings are
in general complex.
On the other hand, ($2n_{g}-1$) phases can be absorbed 
by the field redefinition of $\lmdl{i,0}(x)$ and $\chir{j,0}(x)$.
So the number of the physical phases that cannot be removed is $(n_{g}-1)^{2}$.
In particular, when we consider the case of $n_{g}=3$, 
the following four unremovable CP phases exist.
\begin{eqnarray}
 \eta_{1}&=&\zeta_{11}+\zeta_{22}-\zeta_{12}-\zeta_{21}, \nonumber \\
 \eta_{2}&=&\zeta_{22}+\zeta_{33}-\zeta_{23}-\zeta_{32}, \nonumber \\
 \eta_{3}&=&\zeta_{33}+\zeta_{11}-\zeta_{31}-\zeta_{13}, \nonumber \\
 \eta_{4}&=&\zeta_{12}+\zeta_{23}+\zeta_{31}-\zeta_{21}-\zeta_{13}
 -\zeta_{32}.
\end{eqnarray}
Of course, these phases are independent of the normalization factors of
$\varphi_{\lmdL i,0}(y)$ and $\varphi_{\chiR j,0}(y)$.
Then we will take a quantity:
\begin{equation}
 \Delta\equiv\max(\eta_{1},\eta_{2},\eta_{3},\eta_{4}) \label{Delta}
\end{equation}
as a measure of the CP violation.
When we try to realize the standard model in this framework, 
the Kobayashi-Maskawa (KM) phase is naively thought to be of order 
$\Delta$.

Note that matter fermions $\lmdl{i,0}(x)$ and $\chir{i,0}$ does not 
interact with the NG boson $a_{0}(x)$ directly because only left-handed
(right-handed) zero-mode exists in $\lmd$ ($\chi$).
So no additional CP violating interactions are 
induced from the first and the second terms of Eq.(\ref{Lint1}), 
which play an important role of localizing the zero-modes on our wall.

It can be shown that the CP phase of order one can be realized 
in our mechanism.
As an example, let us consider the case that 
\begin{eqnarray}
 \frac{1}{g^{3}\Lmd}(\hl{1},\hl{2},\hl{3})&=&
 \frac{1}{g^{3}\Lmd}(\hr{1},\hr{2},\hr{3})=(20,12,8), \nonumber \\
 -\frac{1}{g^{2}\Lmd^{2}}(M_{\lmd 1},M_{\lmd 2},M_{\lmd 3})&=&
 \frac{1}{g^{2}\Lmd^{2}}(M_{\chi 1},M_{\chi 2},M_{\chi 3})
 =(-17,0,6). \label{prm-set1}
\end{eqnarray}

\begin{figure}
\leavevmode
\epsfysize=5cm
\centerline{\epsfbox{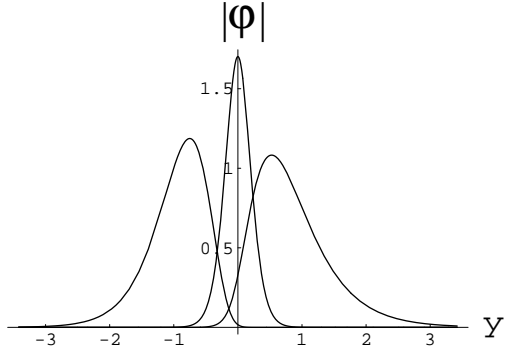}}
\caption{The profiles of the mode functions of the fermions when
the parameters are set as Eq.(\ref{prm-set1}). From left to right,
$\varphi_{\lmdL 1,0}=\varphi_{\chiR 1,0}$, 
$\varphi_{\lmdL 2,0}=\varphi_{\chiR 2,0}$ and 
$\varphi_{\lmdL 3,0}=\varphi_{\chiR 3,0}$ are plotted respectively.
The coordinate $y$ is normalized by $1/(g^{2}\Lmd^{2})$.}
\label{mdfc-in-fat}
\end{figure}

The profiles of the mode functions of fermionic modes are shown 
in Fig.\ref{mdfc-in-fat}.
In this case, the resulting CP phase is $\Delta=0.520$.
Thus we can realize the minimal standard model with the KM phase 
of order one.

In general, additional sources of CP violation appear 
if we try to extend the standard model.
For example, in the supersymmetric standard models, there exist
additional CP phases in the soft SUSY breaking parameters.
If these phases are allowed to be of order one, too large CP violation
might occur\cite{dine-kramer}.
Therefore in such a case, some mechanism is needed 
in order to suppress the CP phases.
We claim that our CP violating mechanism can also realize small CP phases
by setting the parameters: $\{\hl{i}, \hr{j}, M_{\lmd i}, M_{\chi j}\}$,
to different values.
For instance, in the case that
\begin{eqnarray}
 \frac{1}{g^{3}\Lmd}(\hl{1},\hl{2},\hl{3})&=&
 \frac{1}{g^{3}\Lmd}(\hr{1},\hr{2},\hr{3})=(8,8,8), \nonumber \\
 -\frac{1}{g^{2}\Lmd^{2}}(M_{\lmd 1},M_{\lmd 2},M_{\lmd 3})&=&
 \frac{1}{g^{2}\Lmd^{2}}(M_{\chi 1},M_{\chi 2},M_{\chi 3})
 =(-6,0,6),
\end{eqnarray}
the result is $\Delta=3.84\times 10^{-3}$.

There is another way of controlling the CP phases, which seems to be 
somewhat natural, 
especially when the supersymmetric extention of the models are considered.
We will discuss it in the next section.

\section{CP violation due to the coexistence of the walls} \label{coexist}
In this section, we will consider a two-wall system in which
CP is conserved when each wall is isolated, but violated when
the two walls coexist.

Let us consider the following five-dimensional theory.

\begin{equation}
 {\cal L}=-\del^{M}A^{\ast}\del_{M}A-\Lmd^{2}|A|^{2}\left |
 1-\frac{g^{3}}{\Lmd^{3}}A^{3}\right|^{2}, \label{another-model}
\end{equation}
where parameters $\Lmd$ and $g$ have the same mass-dimensions 
as those of the model Eq.(\ref{phi-n-th}) and Eq.(\ref{comp-ex-model}),
and again are real and positive.

\begin{figure}
 \leavevmode
 \epsfysize=5cm
 \centerline{\epsfbox{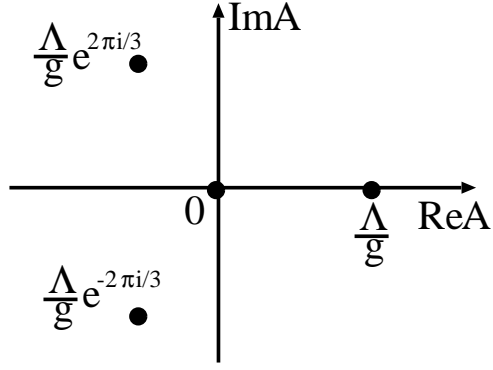}}
 \caption{The degenerate vacua in the theory of Eq.(\ref{another-model}).}
 \label{vacua-plot}
\end{figure}

This model has four degenerate vacua $A=0$, $\Lmd/g$, $(\Lmd/g)e^{2\pi i/3}$ 
and $(\Lmd/g)e^{-2\pi i/3}$, shown in Fig.\ref{vacua-plot}.

There is a domain-wall configuration interpolating the vacua $A=\Lmd/g$ 
and $A=0$ \cite{sakai},
\begin{equation}
 \Acl^{(1)}(y)=\frac{\Lmd}{g\{1+e^{3\Lmd(y-y_{1})}\}^{1/3}}. \label{Acl1}
\end{equation}
Similarly, there is another wall configuration interpolating the vacua $A=0$
and $A=(\Lmd/g)e^{2\pi i/3}$,
\begin{equation}
 \Acl^{(2)}(y)=\frac{\Lmd e^{2\pi i/3}}{g\{1+e^{-3\Lmd(y-y_{2})}\}^{1/3}}. 
 \label{Acl2}
\end{equation}
Here $y_{1}$ and $y_{2}$ roughly represent the position of the walls.

These solutions have definite complex phases that is independent of $y$.
Thus no physical CP phases are induced in the four-dimensional effective 
theory when we live on each of them. 
However if the two walls coexist at finite distance, the background
configuration cannot have a definite phase any longer 
and has a non-trivial phase
that depend on the coordinate of the extra dimension $y$.
Therefore CP is violated in such a situation.

\begin{figure}
 \leavevmode
 \epsfysize=4.5cm
 \centerline{\epsfbox{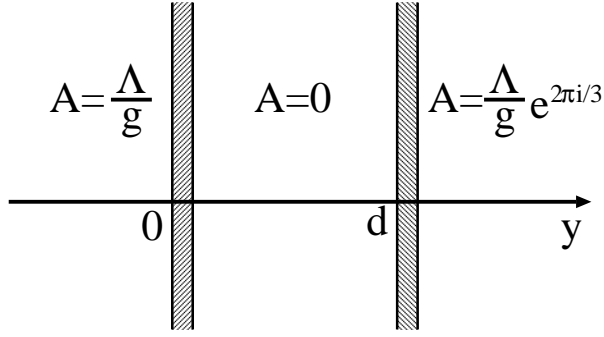}}
 \caption{An interesting double-wall configuration. The wall at $y=0$ is
  $\Acl^{(1)}(y)$ in Eq.(\ref{Acl1}) and the wall at $y=d$ is $\Acl^{(2)}(y)$
  in Eq.(\ref{Acl2}).}
 \label{another-config}
\end{figure}

Now we will investigate the configuration shown in Fig.\ref{another-config}.
We assume that we live on the wall $\Acl^{(1)}(y)$.
From now on, we will set $y_{1},y_{2}=0$ in the definition of $\Acl^{(1)}(y),
\Acl^{(2)}(y)$ of Eq.(\ref{Acl1}),(\ref{Acl2}).

Following Eq.(\ref{Lint1}) in the previous model, we will introduce 
interactions as follows,
\begin{eqnarray}
 {\cal L}_{\rm int}&=&\sum_{i=1}^{n_{g}}\left(
 M_{\lmd i}\bar{\lmd}_{i}\lmd_{i}+\hl{i}(\Re A)\bar{\lmd}_{i}\lmd_{i}
 \right)
 +\sum_{j=1}^{n_{g}}\left(
 M_{\chi j}\bar{\chi}_{j}\chi_{j}-\hr{j}(\Re A)\bar{\chi}_{j}\chi_{j}
 \right) \nonumber \\
 &&-\frac{1}{2}\left(f^{\ast}(A^{\ast})\frac{\del^{2}f}{\del A^{2}}(A)B^{2}
 +h.c.\right)-\left|\frac{\del f}{\del A}(A)B\right|^{2} \nonumber \\
 &&+\left(\sum_{i,j}y_{ij}B\bar{\chi}_{j}\lambda_{i}+h.c.\right),
 \label{Lint2}
\end{eqnarray}
where 
\begin{equation}
 f(A)\equiv \Lmd A\left(1-\frac{g^{3}}{\Lmd^{3}}A^{3}\right),
\end{equation}
and $\hl{i},\hr{j}>0$ and $y_{ij}$ are real.
As in the previous model, 
\begin{displaymath}
 \lmd_{i}=\left(\begin{array}{c} \lmdl{i} \\ \lmdr{i} \end{array}\right)\;, 
 \;\;\;
 \chi_{j}=\left(\begin{array}{c} \chil{j} \\ \chir{j} \end{array}\right)
\end{displaymath}
are five-dimensional Dirac spinor fields and $B$ is a complex scalar field.

Unfortunately, we do not know the exact double-wall configuration
shown in Fig.\ref{another-config}.
However, since the field-configuration of our wall is deformed from
the real configuration $\Acl^{(1)}(y)$ to a complex configuration
by the other wall,
the imaginary part of our-wall configuration can be thought to come from
that of $\Acl^{(2)}(y-d)$, where $d$ is the distance between the walls.
Thus the background configuration $\Acl(y)$ shown in Fig.\ref{another-config} 
is approximated near our wall by
\begin{equation}
 \Acl(y)\simeq\Acl^{(1)}(y)+\Acl^{(2)}(y-d)
 \simeq\Acl^{(1)}(y)+i\Im\Acl^{(2)}(y-d).
 \label{sw-ap}
\end{equation}
The validity of this approximation is confirmed in the appendix.

Then the (pseudo) zero-modes trapped on our wall can be approximated 
near our wall as follows.
\begin{eqnarray}
 \varphi_{\lmdL i}(y)&=&C_{\lmdL i}e^{\int_{0}^{y}\df y'\{\hl{i}\Acl^{(1)}(y')
 +M_{\lmd i}\}}, \nonumber \\
 \varphi_{\chiR j}(y)&=&C_{\chiR j}e^{\int_{0}^{y}\df y'\{\hr{j}\Acl^{(1)}(y')
 -M_{\chi j}\}}, \nonumber \\
 \phi_{B}(y)&=&C_{B}\del_{y}\left(\Acl^{(1)}(y)+i\Im\Acl^{(2)}(y-d)\right), 
 \label{mode-func}
\end{eqnarray}
where $C_{\lmdL i}, C_{\chiR j}$ are complex and $C_{B}$ is real.

\begin{figure}
 \leavevmode
 \epsfysize=5cm
 \centerline{\epsfbox{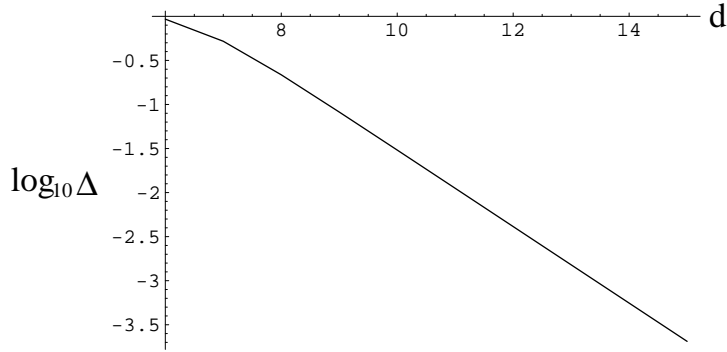}}
 \caption{The measure of the CP violation $\Delta$ as a function of 
  the wall distance in the case of Eq.(\ref{prm-set2}).
  The distance $d$ is normalized by $1/\Lmd$.}
 \label{delta-plot}
\end{figure}

Using these mode functions, effective Yukawa couplings $\yefij$ 
that appear in the four-dimensional effective theory are calculated as
\begin{equation}
 \yefij=y_{ij}\int_{-\infty}^{\infty}\df y\phi_{B}(y)\varphi_{\chiR j}(y)
 \varphi_{\lmdL i}(y).
\end{equation}
The measure of the CP violation $\Delta$, defined in Eq.(\ref{Delta}), 
is calculated from these $\yefij$ 
and it is shown in Fig.\ref{delta-plot} in the case that
\begin{eqnarray}
 \frac{1}{g}(\hl{1},\hl{2},\hl{3})&=&
 \frac{1}{g}(\hr{1},\hr{2},\hr{3})=(20,12,8), \nonumber \\
 \frac{1}{\Lmd}(M_{\lmd 1},M_{\lmd 2},M_{\lmd 3})&=&
 -\frac{1}{\Lmd}(M_{\chi 1},M_{\chi 2},M_{\chi 3})
 =(-1,-4,-7). \label{prm-set2}
\end{eqnarray}
 .

As Fig.\ref{delta-plot} shows, this model can realize a wide range of
the magnitude of $\Delta$, and can easily control it by changing only one
parameter, i.e. the distance between the walls.

When we try to extend the standard model to supersymmetric, 
we should specify the mechanism of the SUSY breaking.
Recently, the author has proposed a new SUSY breaking mechanism
with collaborators
that SUSY is broken due to the coexistence of two different kinds of 
BPS domain walls\cite{maru}.
In this mechanism, we have introduced the other wall in addition to our wall, 
which breaks the supersymmetry preserved by our wall.
Thus if we apply this SUSY breaking mechanism with our CP violating scenario, 
there is a possibility that the other wall is a source of 
not only SUSY breaking but also CP violation.
In such a case, additional CP phases appearing in the soft SUSY breaking
parameters can naturally be suppressed by the wall distance 
just in the same way as those of 
the Yukawa couplings calculated above.

Note that in the localization mechanism we used so far, the mode functions
of the fermionic modes cannot have non-trivial complex phases depending on $y$.
However, this is not an inevitable feature of our scenario.
Indeed, we can make the mode functions of fermions have non-trivial phases
by making the first and the second terms in Eq.(\ref{Lint2}) be 
non-diagonal.
To illustrate this, let us consider the two-generation case.

\begin{equation}
 \cL_{\rm int}=(\bar{\lmd}_{1},\bar{\lmd}_{2})\left(\begin{array}{cc}
 \hl{11}\Re A+M_{\lmd 11} & \hl{12}A+M_{\lmd 12} \\ 
 \hl{12}A^{\ast}+M_{\lmd 12} & \hl{22}\Re A+M_{\lmd 22} \end{array}
 \right)\left(\begin{array}{c} \lmd_{1} \\ \lmd_{2} \end{array}\right),
\end{equation}
where $\hl{ij}$ and $M_{\lmd ij}$ are real.

We define the functions:
\begin{equation}
 \Ml{ij}(y)\equiv -\left(\hl{ij}\Acl(y)+M_{\lmd ij}\right). \;\;\;(i,j=1,2)
\end{equation}
Then the linearized equation of motion for $\lmd_{i}$ is
\begin{equation}
 -i\gamma^{M}\del_{M}\left(\begin{array}{c} \lmd_{1} \\ \lmd_{2} \end{array}
 \right)-\left(\begin{array}{cc} \Re\Ml{11} & \Ml{12} \\
 \Ml{12}^{\ast} & \Re\Ml{22} \end{array}\right)\left(\begin{array}{c}
 \lmd_{1} \\ \lmd_{2} \end{array}\right)=0.
\end{equation}

Defining the operators:
\begin{equation}
 \cO_{\lmdL}\equiv -\del_{y}-\left(\begin{array}{cc} \Re\Ml{11} &
 \Ml{12} \\ \Ml{12}^{\ast} & \Re\Ml{22} \end{array}\right),\;\;\;
 \cO_{\lmdR}\equiv \del_{y}-\left(\begin{array}{cc} \Re\Ml{11} &
 \Ml{12} \\ \Ml{12}^{\ast} & \Re\Ml{22} \end{array}\right),
\end{equation}
the mode functions are defined as solutions of the following equations.
\begin{equation}
 \cO_{\lmdL}\left(\begin{array}{c} \varphi_{{\rm L}1,n}(y) \\
 \varphi_{{\rm L}2,n}(y) \end{array}\right)=m_{n}
 \left(\begin{array}{c} \varphi_{{\rm R}1,n}(y) \\ \varphi_{{\rm R}2,n}(y)
 \end{array}\right), \;\;\;
 \cO_{\lmdR}\left(\begin{array}{c} \varphi_{{\rm R}1,n}(y) \\
 \varphi_{{\rm R}2,n}(y) \end{array}\right)=m_{n}
 \left(\begin{array}{c} \varphi_{{\rm L}1,n}(y) \\ \varphi_{{\rm L}2,n}(y)
 \end{array}\right). \label{mode-eq}
\end{equation}

The bulk fermion fields can be expanded by these mode functions.
\begin{equation}
 \left(\begin{array}{c} \lmd_{{\rm L}1}(x,y) \\ \lmd_{{\rm L}2}(x,y)
 \end{array}\right)=\sum_{n}\lmd_{{\rm L},n}(x)
 \left(\begin{array}{c} \varphi_{{\rm L}1,n}(y) \\ \varphi_{{\rm L}2,n}(y)
 \end{array}\right), \;\;\;
 \left(\begin{array}{c} \lmd_{{\rm R}1}(x,y) \\ \lmd_{{\rm R}2}(x,y)
 \end{array}\right)=\sum_{n}\lmd_{{\rm R},n}(x)
 \left(\begin{array}{c} \varphi_{{\rm R}1,n}(y) \\ \varphi_{{\rm R}2,n}(y)
 \end{array}\right).
\end{equation}

Thus the equation for the zero-modes of $\cO_{\lmdL}$ is
\begin{equation}
 \del_{y}\left(\begin{array}{c} \varphi_{{\rm L}1,0}(y) \\ 
 \varphi_{{\rm L}2,0}(y) \end{array}\right)=-\left(\begin{array}{cc}
 \Re\Ml{11}(y) & \Ml{12}(y) \\ \Ml{12}^{\ast}(y) &
 \Re\Ml{22}(y) \end{array}\right)\left(\begin{array}{c}
 \varphi_{{\rm L}1,0}(y) \\ \varphi_{{\rm L}2,0}(y) \end{array}\right).
 \label{comp-zero-mode}
\end{equation}

Here we denote the asymptotic values of $\Ml{ij}(y)$ as
\begin{equation}
 M_{\pm\infty ij}\equiv\lim_{y\rightarrow\pm\infty}\Ml{ij}(y).
\end{equation}

Then the condition for two zero-modes to exist, that is, for both
solutions of Eq.(\ref{comp-zero-mode}) to decay to zero at infinity, is that
the eigenvalues of the following two matrices are all positive.
\begin{equation}
 \left(\begin{array}{cc} \Re\Minf{11} & \Minf{12} \\
 \Minf{12}^{\ast} & \Re\Minf{22} \end{array}\right),\;\;\;
 -\left(\begin{array}{cc} \Re\Mninf{11} & \Mninf{12} \\
 \Mninf{12}^{\ast} & \Re\Mninf{22} \end{array}\right),
\end{equation}
In other words, 
\begin{equation}
 \left\{\begin{array}{l} \Re\Minf{11}>0 \\
 (\Re\Minf{11})(\Re\Minf{22})-|\Minf{12}|^{2}>0 \end{array}
 \right. ,
\end{equation}
and
\begin{equation}
 \left\{\begin{array}{l} \Re\Mninf{11}<0 \\
 (\Re\Mninf{11})(\Re\Mninf{22})-|\Mninf{12}|^{2}>0 \end{array}
 \right. .
\end{equation}

If this condition is satisfied, two normalizable zero-modes exist
and their asymptotic behaviors are
\begin{eqnarray}
 \left(\begin{array}{c} \varphi_{{\rm L}1,0}(y) \\ \varphi_{{\rm L}2,0}(y)
 \end{array}\right) &\rightarrow &
 \exp\left\{-\left(\begin{array}{cc} \Re\Minf{11} & \Minf{12} 
 \\
 \Minf{12}^{\ast} & \Re\Minf{22} \end{array}\right)y
 \right\}
 \left(\begin{array}{c} C_{1,\infty} \\ C_{2,\infty} \end{array}\right),
 \;\;\;(y\rightarrow\infty)  \\
 &\rightarrow &
 \exp\left\{-\left(\begin{array}{cc} \Re\Mninf{11} & \Mninf{12} \\
 \Mninf{12}^{\ast} & \Re\Mninf{22} \end{array}\right)y
 \right\}
 \left(\begin{array}{c} C_{1,-\infty} \\ C_{2,-\infty} \end{array}\right),
 \;\;\;(y\rightarrow -\infty) \nonumber
\end{eqnarray}
where $C_{i,\pm\infty}$ ($i=1,2$) are complex constants.

Unlike the previous case, these mode functions can have
non-trivial complex phases.
The discussion for the modes in $\chi_{j}$ is the same.

The extention to the three or more generation cases is straightforward.

\section{Summary} \label{summary}
We discussed the origin of the CP violation in the context of 
the brane world scenario,
especially in the case that the three brane where we live is 
a domain wall in a five-dimensional space-time.
In such a case, there is a new possibility that the origin of the CP violation 
in our world is the domain wall itself.

In the large class of models, even if the bulk theory is CP-invariant, 
it has a domain-wall configuration with a non-trivial complex phase 
that depends on the coordinate of the extra dimension $y$.
Then the mode functions of the trapped modes on our wall become 
complex functions.
Since effective couplings in the four-dimensional effective theory are
obtained by overlap integrals of these complex mode functions, non-zero
CP phases appear in the effective theory.
These CP phases cannot be removed by field redefinition because of
the non-trivial $y$-dependence of the background configuration.
As a result, CP violation occurs in the effective theory.

This mechanism can be classified into the spontaneous CP violation, 
but it violates the CP-invariance by the ``complex wall-configuration'',
instead of the complex VEVs of some scalar fields.

In the domain-wall scenario like ours, the hierarchy among 
the Yukawa couplings can easily realized by locating the fermions 
at different positions in our wall\cite{kaplan}.
We showed that our scenario can give $O(1)$ CP phases by using this 
mechanism together.
So we can construct the minimal standard model in our scenario.

When we extend the standard model, additional sources of the CP violation
come out and they often need to be suppressed 
to avoid the contradiction to the experimental data.
Our mechanism can also be applied to such a case because we can 
realize the small CP phases, too.
Especially we considered an interesting case in which the CP violation
in our world is caused by the existence of the other wall, which is located 
distant from our wall along the extra dimension.
When our wall or the other wall is isolated, each wall configuration
has a definite complex phase independent of $y$, and thus CP is not violated
in the effective theory on each wall.
However, when the two walls coexist at finite distance, the background 
configuration no longer has a definite phase and the CP violation occur
on our wall.
We emphasize that our CP violating mechanism does not need any bulk fields
mediating the CP violating effects to our wall or any source of 
the CP violation on the other wall, in contrast to Ref.~\cite{arkani2}.
CP is violated only by the ``coexistence of the walls''.

This double-wall scenario is congenial to the SUSY breaking scenario proposed 
in Ref.~\cite{maru}, in which SUSY is broken due to the coexistence of
two walls.
Namely we can consider an attractive scenario that 
the CP violation and the SUSY breaking
have the common origin, that is, the coexistence of our wall and
the other wall.

One of the characteristic features of our double-wall scenario is
that the CP violation observed on our wall decays exponentially
as the distance between the walls increases.

We also proposed a practical method for estimation of the CP violation 
induced on our wall in the double-wall scenario.
We often encounter the case that only a single-wall configuration is known 
exactly but we do not know an exact double-wall configuration representing 
the coexistence of our wall and the other wall.
We showed that even in such a case, the estimation of the CP violation 
is possible by an appropriate approximation using only a knowledge 
about the single-wall configuration.
The validity of this approximation is also confirmed in the appendix.

Finally, We emphasize that our CP violating mechanism 
has a flexibility to 
realize both types of CP violating models: models with an $O(1)$ CP phase
such as the Kobayashi-Maskawa model and models with small CP phases such as
the supersymmetric standard models.
So the possibility discussed in this paper should be taken 
into account in the model-building of a realistic model 
in the context of the domain-wall scenario.

\begin{center}
{\large\bf Acknowledgements}
\end{center}
The author would like to thank N.Sakai and C.Csaki for useful advice 
and discussions.

\appendix
\section{Calculable double-wall configuration} 
\label{double-wall}
Here we will consider a particular model that has a calculable double-wall 
configuration, and confirm the validity of the approximation Eq.(\ref{sw-ap}).

\subsection{Classical background}
We will introduce a five-dimensional theory as follows.
\begin{equation}
 {\cal L}=-\partial^{M}A^{\ast}\partial_{M}A-\Lmd^{5}\left|\frac{\frac{1}{2}-
 \cos(\frac{g}{\Lmd}A+\alpha)}{2-\cos\frac{g}{\Lmd}A}\right|^{2}, 
 \;\;\;(M=0,1,\cdots,4) 
 \label{our-model}
\end{equation}
where $\alpha$ is a real parameter, and parameters $\Lmd$ and $g$ have
the same mass-dimensions as those of the model Eq.(\ref{phi-n-th}) or
Eq.(\ref{comp-ex-model}) and are positive.
Here, the fifth dimension is compactified on $S^{1}$ whose radius is $R$,
and its coordinate is denoted as $y$, i.e. $y\equiv x^{4}$.
For convenience, we will redefine $A$ and $x^{M}$ as
\begin{equation}
 \frac{g}{\Lmd}A\rightarrow A, \;\;\;
 g\Lmd^{3/2}x^{M}\rightarrow x^{M}, \label{redefine}
\end{equation}
so that $A$ and $x^{M}$ become dimensionless variables.
Then at the classical level, the above theory is equivalent to
the following one.
\begin{equation}
 {\cal L}=-\partial^{M}A^{\ast}\partial_{M}A-\left|\frac{\frac{1}{2}-
 \cos(A+\alpha)}{2-\cos A}\right|^{2}.
 \label{our-model2}
\end{equation}

\begin{figure}
 \leavevmode
 \epsfysize=7cm
 \centerline{\epsfbox{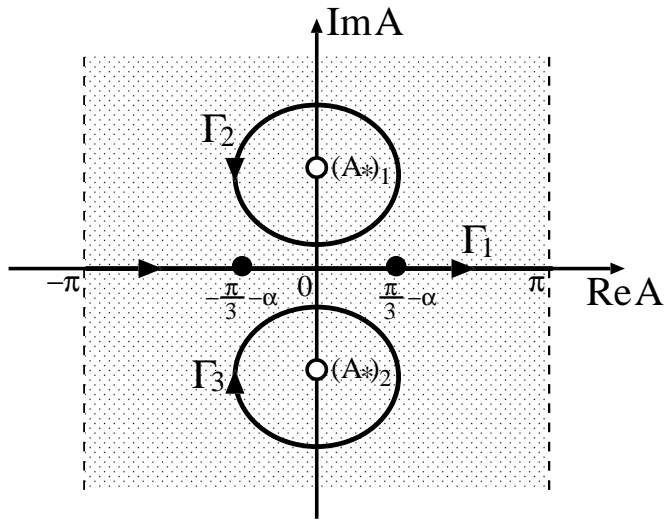}}
 \caption{The target space of the scalar field $A$. The lines $\Re A=\pi$ and
  $\Re A=-\pi$ represent the same line.}
 \label{target-A}
\end{figure}

The target space of the scalar field $A$ has a topology of a cylinder
with two points $(A_{\ast})_{1,2}$ deleted\footnote{
This is similar to the one in Ref.~\cite{hou}.} (Fig.\ref{target-A}),
\begin{equation}
 -\pi\leq\Re A\leq\pi, \;\;\; -\infty < \Im A < \infty,
\end{equation}
and
\begin{equation}
 (A_{\ast})_{1,2}=\pm i \ln(2+\sqrt{3}).
\end{equation}

The model has two vacua at $A=\pm\pi/3-\alpha$ and the potential has
two poles at $A=(A_{\ast})_{1,2}$.
There are three noncontractible cycles, $\Gamma_{1}, \Gamma_{2}$ and 
$\Gamma_{3}$ depicted in Fig.\ref{target-A}.

Now let us consider a vacuum configuration that depends only on $y$.
We will seek a domain-wall configuration that winds around the pole 
$(A_{\ast})_{1}$ counterclockwise as $y$ increases.
Its trajectory on the target space corresponds to the cycle $\Gamma_{2}$.
Such a configuration is topologically stable.
To obtain such a configuration, we will dimensionally reduce our model
to the four-dimensional theory in terms of, for example, $x^{3}$-direction.
Then the problem is reduced to seeking a domain-wall configuration 
in the four-dimensional theory.

In this case, our model Eq.(\ref{our-model2}) can be regarded as a bosonic
part of a four-dimensional ${\cal N}=1$ supersymmetric model,
\begin{equation}
 {\cal L}^{(4)}=\bar{\Phi}\Phi|_{\theta^{2}\bar{\theta}^{2}}
 +W(\Phi)|_{\theta^{2}}+h.c.,
\end{equation}
where $\Phi=A+\sqrt{2}\theta\psi+\theta^{2}F$ is a chiral superfield,
and $W(\Phi)$ is a superpotential as follows.
\begin{equation}
 W(\Phi)=\cos\alpha\cdot \Phi+(\frac{1}{2}-2\cos\alpha)\frac{2}{\sqrt{3}}
 \tan^{-1}(\sqrt{3}\tan\frac{\Phi}{2})+\sin\alpha\cdot\ln(2-\cos\Phi).
\end{equation}

We will seek a BPS configuration in this model.
Such a configuration can be found by using a method proposed in 
Ref.~\cite{hou}.

The period corresponding to the cycle $\Gamma_{2}$ is 
\begin{eqnarray}
 \Delta W&=&\oint_{\Gamma_{2}}\frac{\partial W}{\partial A}\df A
 =\oint_{\Gamma_{2}}\frac{\frac{1}{2}-\cos(A+\alpha)}{2-\cos A}{\rm d}A 
 \nonumber\\
 &=&\frac{2\pi}{\sqrt{3}}\left\{2\cos\alpha-\frac{1}{2}-i\sqrt{3}\sin\alpha
 \right\} \label{period-gm2}
\end{eqnarray}

Thus the BPS equations are
\begin{equation}
 \frac{\df A}{\df y}=e^{i\delta}\frac{\del W^{\ast}}{\del A^{\ast}}\;,\;\;\;
 \frac{\df A^{\ast}}{\df y}=e^{-i\delta}\frac{\del W}{\del A}, \label{bps-eq}
\end{equation}
where $\delta\equiv\arg(\Delta W)$.

\begin{figure}
 \leavevmode
 \epsfysize=4.5cm
 \centerline{\epsfbox{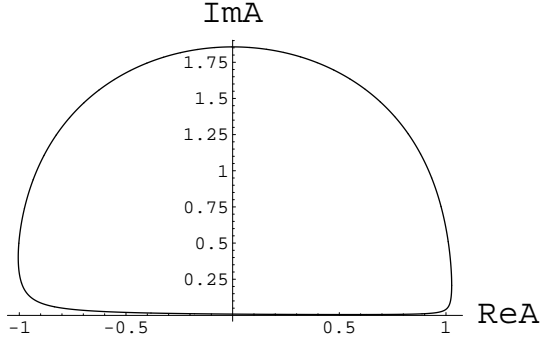}}
 \caption{the BPS trajectory homotopical to the cycle $\Gamma_{2}$.
  This is the contour corresponding to $\alpha=0.01$ ($I_{\infty}=-0.00357$)
  and $I_{0}=-0.00448$.}
 \label{closed-path}
\end{figure}

Here we will introduce the multivalued function $I(A,A^{\ast})$, which is 
defined as
\begin{equation}
 I=\Im(e^{-i\delta}W).
\end{equation}
This is the integral of motion of Eq.(\ref{bps-eq}).
Note that a trajectory of a BPS configuration on the field space is 
a contour line of $I(A,A^{\ast})=I_{0}$ where $I_{0}$ is a real constant.
Here we are interested in the field configuration that has a wall structure,
so we will consider a contour line that passes near the two vacua 
$A=\pm\pi/3-\alpha$ as shown in Fig.\ref{closed-path}.
It can be obtained by putting $I_{0}$ close to the value
$\Iinf\equiv\Im(e^{-i\delta}W(\pi/3-|\alpha|))$ from below.

To parametrize the contour shown in Fig.\ref{closed-path},
we will introduce $\theta\equiv\arg\{i(A-(A_{\ast})_{1})\}$ as a parameter,
where $A$ is a point on the contour.
The relation between $\theta$ and $y$ is obtained from Eq.(\ref{bps-eq}),
\begin{equation}
 y(\theta)=\int_{0}^{\theta}\df\theta\frac{\df A}{\df\theta}e^{-i\delta}
 \frac{2-\cos A^{\ast}}{\frac{1}{2}-\cos(A^{\ast}+\alpha)}.
\end{equation}
Here we set the initial condition as $\theta=0$ at $y=0$.
By using this relation, we can obtain the classical solution $\Acl(y)$
for each value of $I_{0}$.

\begin{figure}
 \leavevmode
 \epsfysize=8cm
 \centerline{\epsfbox{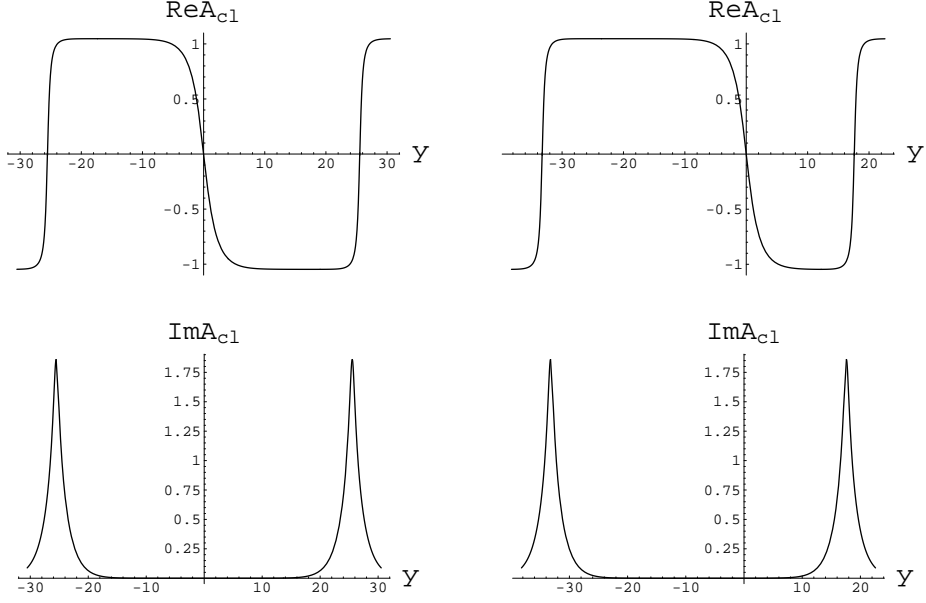}}
 \caption{The profile of the classical configuration $\Acl(y)$ in the case
  of $\alpha=0$ (the left plots) and $\alpha\neq 0$ (the right plots).}
 \label{Acl-profile}
\end{figure}

At first, let us consider the case of $\alpha=0$.
In this case, $\Iinf=0$ and the configuration becomes two BPS domain walls.
The distance between them goes to infinity in the limit of 
$I_{0}\rightarrow\Iinf$.
These two domain walls preserve the same supersymmetry in contrast to
the case discussed in Ref.~\cite{maru}.
This configuration has an equidistant-wall structure shown by the left plots 
in Fig.\ref{Acl-profile}.
Here the period of the configuration is set to be $2\pi R$
in order to realize a double-wall system.

The wall located at $y=0$ becomes a real configuration $\Acl^{(1)}(y)$
in the limit of $R\rightarrow\infty$, 
and we will regard it as our wall in the following.
The inverse function of $\Acl^{(1)}(y)$ can be calculated analytically as
\begin{equation}
 y=\Acl^{(1)}-2\sqrt{3}\tanh^{-1}\left(\sqrt{3}\tan\frac{\Acl^{(1)}}{2}
 \right).
\end{equation}
On the other hand, the other wall located at $y=\pi R$ is a complex 
configuration even if $R$ goes to infinity.
We will denote this wall in the limit of $R\rightarrow\infty$ 
as $\Acl^{(2)}(y)$.

Next we will consider the case of $\alpha\neq 0$.
In this case, when $I_{0}$ is put close to $\Iinf$, the contour approaches
the two vacua in an asymmetric manner and thus the configuration has 
a non-equidistant-wall structure shown by the right plots 
in Fig.\ref{Acl-profile}, 
in contrast to the cases in Ref.~\cite{hofmann,hou}.
Unlike the previous case, our wall does not become a real configuration
even in the limit of $R\rightarrow\infty$ (i.e., $I_{0}\rightarrow\Iinf$) and
becomes a structure such that two BPS domain walls are finitely separated in
a non-compact space.
This configuration is similar to the one in Ref.~\cite{troitsky}.

From these facts, for a given compactified radius $R$, 
the distance between our wall and the other wall can be set to an arbitrary 
value by adjusting the constant $I_{0}$ and the parameter $\alpha$~\footnote{
Strictly speaking, there is a lower bound for the wall distance.}.
For simplicity, however, we will limit ourselves to the case of 
$\alpha=0$ in the following discussion.

Although the classical configuration $\Acl(y)$ is obtained in 
the four-dimensional ${\cal N}=$~$1$ supersymmetric model, 
this configuration 
can be regarded as a domain wall 
in the five-dimensional non-supersymmetric theory of Eq.(\ref{our-model2}), 
because all we used in 
the above derivation is a bosonic part of the theory.
Thus in the following, we will regard $\Acl(y)$ as a desired classical 
double-wall configuration of the model of Eq.(\ref{our-model2}).

\subsection{Estimation of CP violation}
We will follow the procedure in Section~\ref{est-ob-cp}
to estimate the CP phase in the four-dimensional effective theory.
At first, we will add the following interaction terms 
to the original Lagrangian Eq.(\ref{our-model2}),
\begin{eqnarray}
 \cL_{\rm int}&=&\sum_{i=1}^{n_{g}}\left(M_{\lmd i}\bar{\lmd}_{i}\lmd_{i}
 +\hl{i}(\Re A)\bar{\lmd}_{i}\lmd_{i}\right)
 +\sum_{j=1}^{n_{g}}\left(M_{\chi j}\bar{\chi}_{j}\chi_{j}
 -\hr{j}(\Re A)\bar{\chi}_{j}\chi_{j}\right) \nonumber \\
 &&-\frac{1}{2}
 \left(f^{\ast}(A^{\ast})\frac{\del^{2}f}{\del A^{2}}(A)B^{2}+h.c.\right)
 -\left|\frac{\del f}{\del A}(A)B\right|^{2} \nonumber \\
 &&+\left(\sum_{i,j}y_{ij}B\bar{\chi_{j}}\lambda_{i}+h.c.\right),
 \label{Lint3}
\end{eqnarray}
where 
\begin{equation}
 f(A)\equiv\frac{\frac{1}{2}-\cos A}{2-\cos A}.
\end{equation}
Similarly to Section~\ref{est-ob-cp}, $\hl{i},\hr{j}>0$ and 
$M_{\lmd i}$, $M_{\chi j}$ and $y_{ij}$ are real.

Then the mode functions of the zero-modes $\lmdl{i,0}(x)$, $\chir{j,0}(x)$
and $b_{0}(x)$ in $\lmd_{i}$, $\chi_{j}$ and $B$, which are localized on
our wall, are
\begin{eqnarray}
 \varphi_{\lmdL i,0}(y)&=&C_{\lmdL i}e^{\int_{0}^{y}\df y'
 (\hl{i}\Re\Acl(y')+M_{\lmd i})},
 \nonumber \\
 \varphi_{\chiR j,0}(y)&=&C_{\chiR j}e^{\int_{0}^{y}\df y'
 (\hr{j}\Re\Acl(y')-M_{\chi j})},
 \nonumber \\
 \phi_{B,0}(y)&=&C_{A}\del_{y}\Acl(y), \label{pNG2}
\end{eqnarray}
where $C_{\lmdL i}, C_{\chiR j}$ are complex and $C_{A}$ is real.

Strictly speaking, we must check up whether the above functions are periodic
because $y$ is the coordinate of the extra dimension compactified on
$S^{1}$ in this model.
This condition is not satisfied unless 
$\int_{0}^{2\pi R}{\rm d}y'(\hl{i}\Re\Acl(y')+M_{\lmd i})=
\int_{0}^{2\pi R}{\rm d}y'(\hr{j}\Re\Acl(y')-M_{\chi j})=0$.
Thus there is no fermionic zero-mode in the strict meaning.
Nevertheless, since there exist the zero-modes in a single-wall
background\cite{jackiw}, it is natural to suppose that ``pseudo-zero-modes''
exist when the distance between the walls is large enough.
So we will assume that there exist the pseudo-zero-modes in $\lmd_{i}$ and
$\chi_{j}$, and their mode functions are well approximated 
by Eq.(\ref{pNG2}) near our wall.

The effective Yukawa couplings $\yefij$ involving $\lmdl{i,0}$, $\chir{j,0}$
and $b_{0}$ are calculated as
\begin{equation}
 \yefij=y_{ij}\int_{-\pi R}^{\pi R}\df y\phi_{B,0}(y)\varphi_{\chiR j,0}(y)
 \varphi_{\lmdL i,0}(y). \label{yeff2}
\end{equation}

\begin{figure}
\leavevmode
\epsfysize=5cm
\centerline{\epsfbox{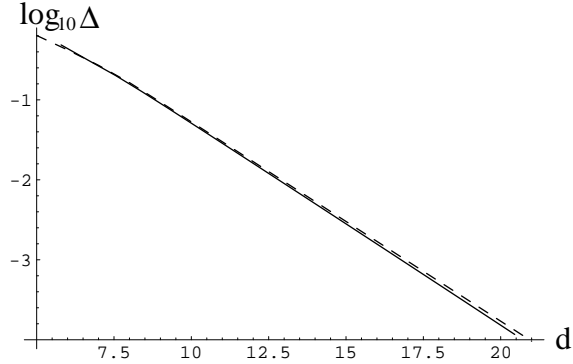}}
\caption{The measure of the CP violation $\Delta$ as a function of 
  the wall distance in the case of Eq.(\ref{prm-set3}). The solid
  line represents the result by using the exact configuration $\Acl(y)$
  and the dashed line is the result by the approximation Eq.(\ref{sw-ap2}).
  The distance $d$ is normalized by $1/(g\Lmd^{3/2})$.}
\label{delta-plot2}
\end{figure}

Then we can calculate the quantity $\Delta$ defined by Eq.(\ref{Delta}),
from these $\yefij$.
It is shown by the solid line in Fig.\ref{delta-plot2} 
in the case that\footnote{
By considering the redefinition Eq.(\ref{redefine}), 
we have restored the dependence of $g$ and $\Lmd$.}
\begin{eqnarray}
 \frac{1}{g^{2}\Lmd^{1/2}}(\hl{1},\hl{2},\hl{3})&=&
 \frac{1}{g^{2}\Lmd^{1/2}}(\hr{1},\hr{2},\hr{3})=(20,12,8), \nonumber \\
 \frac{1}{g\Lmd^{3/2}}(M_{\lmd 1},M_{\lmd 2},M_{\lmd 3})&=&
 -\frac{1}{g\Lmd^{3/2}}(M_{\chi 1},M_{\chi 2},M_{\chi 3})=(-16,0,6).
 \label{prm-set3}
\end{eqnarray}
As we can see from this plot, the CP violating effects decay 
exponentially as the wall distance increases.

Next we will confirm the validity of the approximation Eq.(\ref{sw-ap}).
We approximate $\Acl(y)$ near our wall by
\begin{equation}
 \Acl(y)\simeq\Acl^{(1)}(y)+i\Im\Acl^{(2)}(y-d)+i\Im\Acl^{(2)}(y+d),
 \label{sw-ap2}
\end{equation}
where $d=\pi R$ is the distance between the walls.
Here note that there are both contributions of the other wall 
at $y=d$ and $y=-d$.

Namely, (pseudo-)zero-modes trapped on our wall can be approximated
near our wall as follows.
\begin{eqnarray}
 \varphi_{\lmdL i,0}(y)&=&C_{\lmdL i}e^{\int_{0}^{y}\df y'
 (\hl{i}\Re\Acl^{(1)}(y')+M_{\lmd i})},
 \nonumber \\
 \varphi_{\chiR j,0}(y)&=&C_{\chiR j}e^{\int_{0}^{y}\df y'
 (\hr{j}\Re\Acl^{(1)}(y')-M_{\chi j})},
 \nonumber \\
 \phi_{B,0}(y)&=&C_{A}\del_{y}\left(\Acl^{(1)}(y)+i\Im\Acl^{(2)}(y-d)
 +i\Im\Acl^{(2)}(y+d)\right), 
\end{eqnarray}

The measure of the CP violation $\Delta$ calculated by these
approximate mode functions is plotted by the dashed line
in Fig.\ref{delta-plot2}.

As shown in Fig.\ref{delta-plot2}, we can conclude that
the approximation Eq.(\ref{sw-ap2}) is pretty good.

\end{document}